\documentclass[graybox]{svmult}

\usepackage{mathptmx}       
\usepackage{helvet}         
\usepackage{courier}        
\usepackage{type1cm}        
\usepackage{makeidx}         
\usepackage{graphicx}        
\usepackage{multicol}        
\usepackage[bottom]{footmisc}

\usepackage{natbib}
\usepackage{amssymb}

\makeindex             

\begin{document}

\title*{X-ray Pulsars}
\author{Roland Walter \& Carlo Ferrigno}
\institute{ISDC, Geneva Observatory, University of Geneva, Chemin d'Ecogia 16, CH-1290 Versoix, Switzerland, \email{roland.walter@unige.ch}
}

\maketitle

\abstract{X-ray pulsars shine thanks to the conversion of the gravitational energy of accreted material to X-ray radiation. The accretion rate is modulated by geometrical and hydrodynamical effects in the stellar wind of the pulsar companions and/or by instabilities in accretion discs. Wind driven flows are highly unstable close to neutron stars and responsible for X-ray variability by factors $\gtrsim 10^3$ on time scale of hours. Disk driven flows feature slower state transitions and quasi periodic oscillations related to orbital motion and precession or resonance. On shorter time scales, and closer to the surface of the neutron star,  X-ray variability is dominated by the interactions of the accreting flow with the spinning magnetosphere. When the pulsar magnetic field is large, the flow is confined in a relatively narrow accretion column, whose geometrical properties drive the observed X-ray emission. In low magnetized systems, an increasing accretion rate allows the ignition of powerful explosive thermonuclear burning at the neutron star surface. Transitions from rotation- to accretion-powered activity has been observed in rare cases and proved the link between these classes of pulsars.
}

\section{Introduction}

Accreting binaries \index{accreting binaries} (Casares and Israelian 2016) are among the brightest X-ray point sources of our Galaxy and the first ones to be detected in the early days of X-ray\index{X-ray} astronomy \citep{1964Natur.204..981G}. Their X-ray emission originates from the dissipation of gravitational energy\index{gravitational energy} in material accreted from a companion star to a compact object. A large fraction of the brightest X-ray binaries harbour neutron stars, known as``accreting pulsars\index{accreting pulsars}" or ``X-ray pulsars\index{X-ray pulsars}".

Neutron stars\index{neutron stars} are the remnants of supernova\index{supernova} explosion and are unique laboratories for the study of extreme densities, momentum, gravity and magnetic fields. Understanding them requires all fields of modern physics: plasma physics, electrodynamics, magneto-hydrodynamics, general relativity and quantum physics  (Geppert, 2016; Weber, 2016; Wex, 2016). X-ray binaries\index{X-ray binaries} are the result of complex evolutionary scenarios (van den Heuvel, 2016) established using the full arsenal of stellar evolution, supernova explosions, exchange and accretion processes.

The neutron star magnetic field\index{magnetic field} (Zhang, 2016) and surface play key roles to determine how matter is accreted and where energy is dissipated. The pulsar magnetosphere\index{magnetosphere} dominates the flow of the accreted material within the Alfv\'en surface\index{Alfv\'en surface}, where the bulk kinetic energy density of the gas is comparable to that of the magnetic field. This surface depends on the geometry of the magnetic field, of the accreting flow\index{accreting flow} and of their interaction. Its characteristic size 
is a hundred times larger than the neutron star for a magnetic field of $10^{12}$ G and a luminosity reaching a fraction of the Eddington limit, and can reach the surface of the neutron star when the magnetic field is below $10^8$ G.
An offset between rotation and magnetic axes is required to obtain the flux modulation characterising bright X-ray pulsars.

High ($10^{11-14}$ G) surface magnetic fields are detected in young ($10^6$ y) high mass X-ray binaries\index{high mass X-ray binaries (HXMB)} (HMXB) with massive, O \& B type, stellar companions \citep[but see the peculiar Her~X-1 with a magnetic field of $10^{12}$\,G and a 2 M$_{\odot}$ companion;][]{1978ApJ...219L.105T}. Low surface magnetic fields ($\lesssim 10^{10}$ G) are present in old binary systems with low mass ($\lesssim 1$ M$_{\odot}$) companions (LMXB). HMXBs are  concentrated in the Galactic arms, close to their birthplace. LMXB populate the bulge of the Galaxy and globular clusters, where they can also form through stellar capture.

The Milky Way contains about 130 and 180 bright ($>10^{35}$ erg/s) high and low mass X-ray binaries\index{low mass X-ray binaries (LMXB)}, respectively \citep{2015A&ARv..23....2W,2007A&A...469..807L}. The brightest sources dominate the X-ray emission of the Galaxy at a level of $\sim 10^{38}$ and $\sim 10^{39}$ erg\,s$^{-1}$ for the high and low mass systems. 
HMXBs and the hot interstellar gas dominate the X-ray luminosity of star forming galaxies, a tracer of their stellar formation rate \citep{2003MNRAS.339..793G,2012MNRAS.419.2095M}. 

In this chapter, we will concentrate on the accretion flows driving X-ray variability (section 2 \& 3) and on the mechanisms driving X-ray emission in the direct vicinity of the pulsars (section 4 \& 5).

\section{Wind driven flows}

In HMXBs, the pulsar attracts a small fraction of the stellar wind of its companion \citep{1944MNRAS.104..273B,1973ApJ...179..585D}. In classical wind accreting systems, Bondi-Hoyle accretion\index{Bondi-Hoyle accretion} takes place along the neutron star orbit and the accretion rate remains usually low. High accretion rates are expected in close systems, where the companion is practically filling its Roche lobe. The wind is then focussed through a tidal stream\index{tidal stream} and, if its angular momentum is large enough, a transient accretion disk\index{accretion disk} structure may form. Roche-lobe overflow\index{Roche-lobe overflow} from a high-mass companion is rarely observed, as the compact object is quickly enshrouded by the atmosphere of its companion.

Flares reaching the Eddington luminosity, occur when the compact object crosses a dense component of the stellar wind, usually expelled by a fast rotating main sequence star, featuring emission lines in the optical band. These systems are identified as ``Be X-ray binaries\index{Be X-ray binaries}''.

\subsection{Classical systems}

The instantaneous X-ray luminosity of an accreting pulsar with moderate magnetic field ($\sim 10^{12}$\,G) in a HMXB system is mostly determined by the density and velocity of the stellar wind close to the compact object.
The amplitude of the X-ray variability is determined by the pulsar orbital eccentricity\index{eccentricity}, clumping\index{clumping} and variability of the stellar wind, and by hydrodynamical effects induced by the gravity and photo-ionisation of the neutron star.

\begin{figure}
\centerline{\includegraphics[width=8cm]{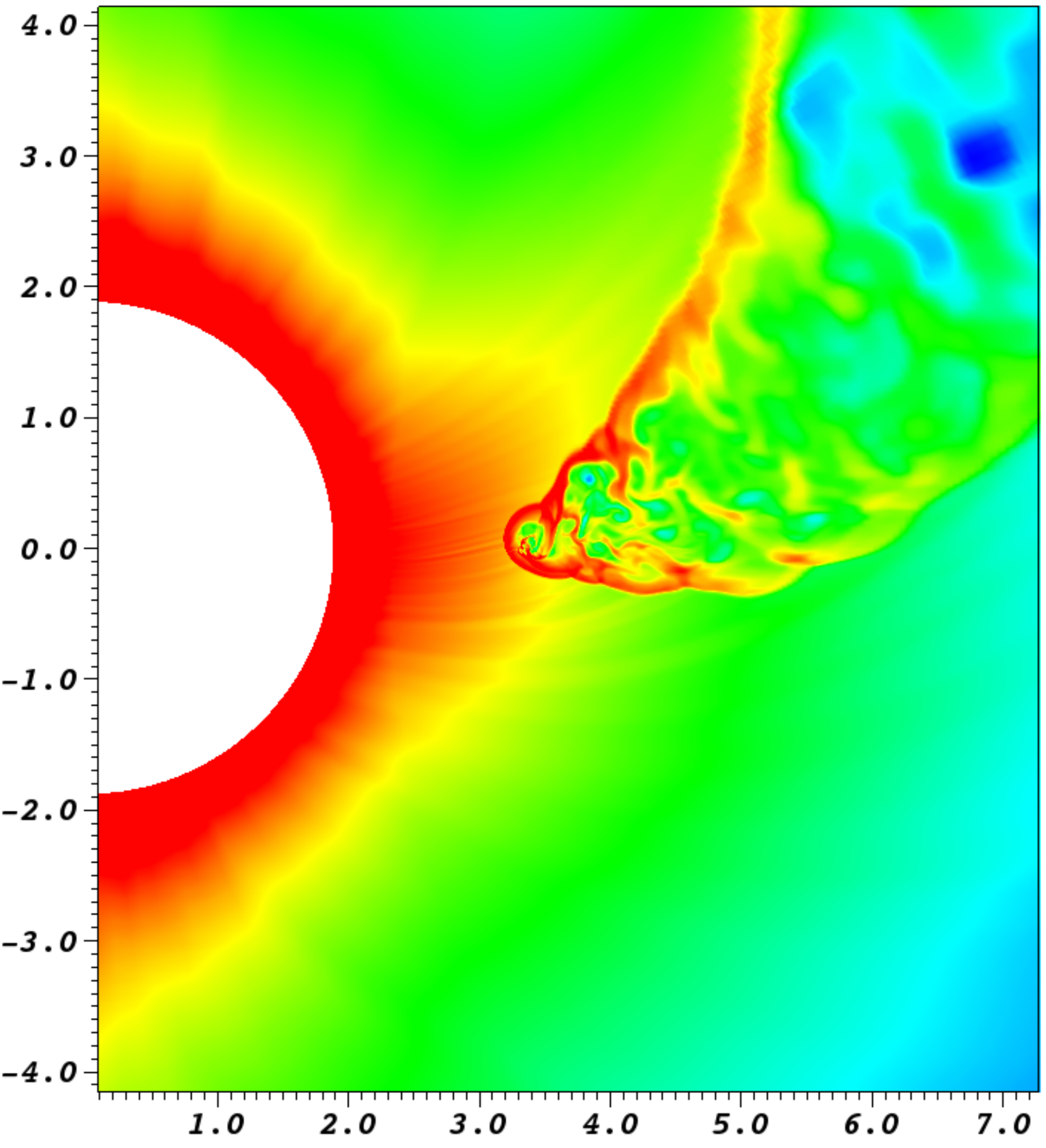}}
\caption{Wind flow in the classical system Vela X-1. The wind acceleration parameters were adjusted to obtain a perfect agreement with the X-ray flux observed on average before, during and after the neutron star eclipse by the companion. Axis are labelled in units of $10^{12}$\,cm. The color scale maps the density from  $10^{-15}$ (blue) to $10^{-13}$ (red) g\,cm$^{-3}$ \citep[simulation from][]{2015A&A...584A..25M}}
\label{fig:wind}
\end{figure}

The variability of the accretion rate by a factor of 10-100 in wind-fed systems in circular orbits was successfully explained by hydrodynamical simulations\index{hydrodynamics} \citep{1990ApJ...356..591B}. \cite{2015A&A...575A..58M} have included the effect of photo-ionisation on the wind acceleration and could probe the dynamic of the region surrounding the neutron star and, in particular, the collision between the primary stellar wind, slowed down by photo-ionisation and a gas stream flowing back inwards from above the neutron star.  As shown in Fig.~\ref{fig:wind}, a shock\index{shock} front is generated, moving inwards and outwards regularly and creating low density bubbles, i.e., periods of very low X-ray luminosity. This generates instantaneous accretion rates varying by $10^3$, and transient modulations similar to these observed in Vela X-1 \citep{2008A&A...492..511K}. This back and forth shock motion occurs high above the magnetosphere and can be amplified further by an induced change of geometry of the accretion column\index{accretion column}. 

\cite{2013MNRAS.428..670S} have shown that two regimes of subsonic accretion are possible at the boundary of the magnetosphere depending on whether or not the plasma is cooled by Compton processes (high vs low accretion rate). At lower luminosity, X-ray photons are emitted perpendicular to the neutron star surface, inverse Compton cooling is less efficient and a change of the X-ray spin modulation of the light curve is expected \citep{2011A&A...529A..52D}. 

\cite{2007AstL...33..149G} suggested that high variability factors could be generated by Kelvin-Helmoltz instability at the magnetospheric boundary, leading to a magnetic gating\index{magnetic gating} of the accretion flow. This requires large magnetic fields ($> 10^{13}$ G) which are contrasting with the observations \citep{2015MNRAS.447.2274B}. It is unclear if magnetic gating is at play in HMXBs.

The comparison of hydrodynamical simulations and X-ray observations allow to probe the stellar wind velocity and density fields \citep{2015A&A...584A..25M}. GX~301$-$2 and OAO~1657$-$415 feature peculiar variability patterns that could be related to the accretion of dense streams and large scale structures in the wind of their massive companions. Long term modulation of the accretion rate along the orbit because of eccentricity is observed in addition in several systems. 

\subsection{Systems close to Roche-lobe overflow}

When the companion star in a HMXB gets closer to Roche-lobe overflow, a tidal stream develops, focuses the wind and increases the wind density close to the compact object and the X-ray luminosity \citep{1991ApJ...371..684B}. Enhanced obscuration by the stream trailing the neutron star is observed first at late orbital phases and covering more and more of the orbit when the neutron star gravity becomes dominant. Once the companion is close to overflowing its Roche lobe, deep spiral-in is unavoidable \citep{1973A&A....25..387V} and results in a common envelope\index{common envelope} phase \citep{2000ARA&A..38..113T}.

Five super-giant HMXBs feature persistently high obscuration (N$_H>10^{23}$ cm$^{-2}$) and  short orbital periods. They all reach X-ray high luminosities $>10^{36}$ erg/s. Two obscured systems have longer orbital periods ($\approx 10$ days) and in these cases the obscuration is probably driven by unusually low wind velocity or by the environment. It is plausible therefore to assume that obscured sgHMXBs are classical systems in transition to Roche lobe overflow, or with relatively low velocity winds. As neutron stars can cut-off wind acceleration via ionisation \citep{1990ApJ...365..321S}, the wind of their companions can be slower on average than in isolated stars. 

Super-giant fast X-ray transients\index{Super-giant fast X-ray transients} were identified as a new class of sources. These hard X-ray transients produce short and bright flares with typical durations of a few ksec. Further analysis indicates that many of them could be interpreted as classical or eccentric systems \citep{2015A&ARv..23....2W}. Four of them are really peculiar: they have short orbital periods (3-6 days), so should be close to Roche lobe overflow, but feature anomalously low luminosities ($<10^{34}$ erg/s), excepting during episodic flares. Abnormally low mass-loss rates or high wind velocities could explain low accretion, but not high variability. Extreme wind clumping seems unlikely very close to the stellar surface. As none of these four sources show pulsation, it is possible that the magnetic and rotation axis are close together and that X-rays are beamed \citep{1973A&A....25..233G} in directions most of the time not favourable for the observer. Flares could be related to periods when the geometry of the accretion column changes significantly.

X-ray beaming\index{beaming} has also been used to explain luminosities reaching $10^{40}$ erg/s (i.e. 100 times Eddington) observed in the ultra luminous X-ray source\index{ultra luminous X-ray source} M82 X-2, an X-ray pulsar in a system close to Roche lobe overflow \citep{2014Natur.514..202B} or of Be nature. Magnetic gating is an alternate explanation both for M82 X-2 \citep{2015MNRAS.454.2539M,2016MNRAS.461....2P} and for the four SFXTs above \citep{2008ApJ...683.1031B} but would require high magnetic fields for which we have no evidence otherwise.

\subsection{Be X-ray binaries\index{Be X-ray binaries}}

Besides the persistent accreting sources described above, transient systems with Be stars as secondaries constitute a substantial part of all HMXBs. Be stars are non super-giant B-type stars that have shown emission lines in their spectra, originating from a circumstellar disk expelled irregularly by a rapidly rotating star \citep{2003PASP..115.1153P}. These equatorial mass ejection (which are independent of binarity) produce a dense ring of gas around the star and a shallow polar wind. The disk gives rise to a sudden appearance of hydrogen emission lines in their optical spectrum. Some Be system seems permanently active, others suddenly enter the Be phase for weeks to years \citep{1987pbsp.book.....S}.

When a neutron star orbits close enough to a Be companion, mass ejection could be accompanied by accretion and by an X-ray outburst. Two types of outbursts are observed: type I outbursts are caused by enhanced mass accretion rate close to periastron, last for 0.2-0.3 P$_{\rm orb}$ and peak to $\approx 10^{37}$ erg\,s$^{-1}$; the rare type II outbursts, reaching the Eddington luminosity can last for several orbital periods. Low eccentricity Be systems are more efficient X-ray emitters. High eccentricity systems eventually circularise because of tidal effects.

\section{Disk driven accretion flows}
\label{sec:lmxb}
\begin{figure}
\centerline{\includegraphics[width=8cm]{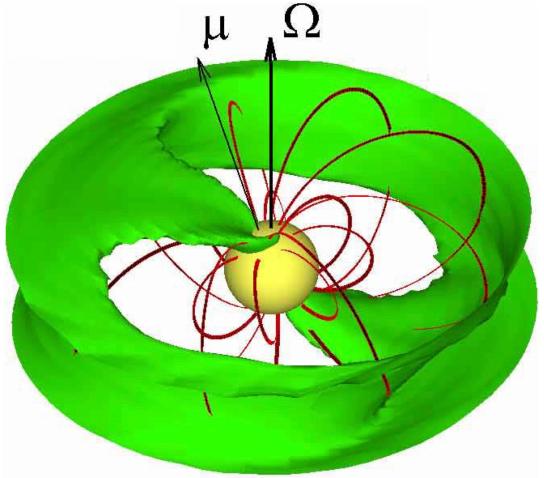}}
\caption{Magneto-hydrodynamical simulation of disk-mediated accretion onto a neutron star with a low magnetic field ($B\sim10^8$\,G). The vectors $\Omega$ and $\mu$ represent the spin and the dipole magnetic field; magnetic field lines are represented in red, while a representative surface of equal density is plotted in green  \citep[the typical case of an accreting millisecond pulsar; from][]{Kulkarni2008}.}
\label{fig:flow}
\end{figure}

Accretion discs\index{accretion disc} form when the stream of material flowing from the companion star intersects with itself before reaching the magnetosphere of the X-ray pulsar. This happens in LMXBs, and sometimes in HMXBs when a tidal stream develops and dominates Bondi-Hoyle accretion. Magneto-hydrodynamical simulations have investigated the transition of the flow from the disc to the central object, where the stream impinges the surface at the magnetic poles (see an example in Fig.~\ref{fig:flow}).

The angular momentum carried from the companion star through the accretion disk can be transferred to the neutron star at the magnetospheric boundary. The inner disk is forced to co-rotate with the neutron star before being channelled to the accretion column and gradually spins up or down the pulsar until an equilibrium period is reached. The interaction between the disk and the magnetosphere is not yet well understood. The differential rotation between the disk and the dipolar magnetosphere generates a toroidal magnetic field \citep{1979ApJ...234..296G} and the plasma leaves the disk towards the dipolar field lines. This model is roughly consistent with the accretion torque\index{accretion torque} behaviour observed during Be/X-ray pulsar flares \citep{1997ApJS..113..367B} when the accretion rate changes dramatically. If the combination of the magnetic field and neutron star rotation is too strong, as compared to the infalling plasma pressure, it can prevent accretion through propeller effects \citep{1970Schwarzman,1975A&A....39..185I}. This happens in the initial phases of pulsar evolution and when the accretion rate lowers, effectively quenching accretion.


The light curves of X-ray pulsars might be modulated periodically by quasi periodic oscillations \index{quasi periodic oscillation} (QPOs): a variability pattern, which is normally transient and with a frequency slightly oscillating around a central value. The high-frequency (kHz) QPOs observed in LMXBs are consistent with orbital motions at the boundary of the magnetosphere \citep{2000ARA&A..38..717V}. In some cases, their frequency is so high that part of the accretion stream should extend further in the magnetosphere \citep{1998ApJ...508..791M,1999ApJ...520..763P}. The accretion flow undergoes a discontinuous change, when the QPO frequency exceeds or falls below the pulsar spin, the disc boundary extends inside or outside the radius at which the pulsar's magnetic field co-rotates with the Keplerian flow, and when accretion is favoured or suppressed, respectively \citep{Bult2015}. Twin kHz QPO can form and might indicate resonances in the disk \citep{2003astro.ph..8179L}. Low frequency QPO (0.1-100 Hz) are probably the signature of disk precession\index{precession} because of frame dragging\index{frame dragging} \citep{1999NuPhS..69..135S} and of accretion instabilities\index{accretion instabilities}.

Discs in HMXBs are unstable because of the hydrodynamical and 3D nature of the wind that feeds them. They are short lived and could rotate in alternate directions. Discs in LMXBs are more stable geometrically, but develop intrinsic instabilities which are likely related to temperature dependent viscosity. When the temperature increases, the mass accretion rate increases, the disc gets hotter, and material falls towards the compact object in a run-away reducing the surface density and returning back to a cooler state.  Recurrent episodes of accretion outbursts may occur triggered by this mechanism, each one lasting from weeks to months, while quiescence can last many years. As X-ray pulsars are luminous objects, the irradiation and heating of the accretion disk by the central source launches thermally driven winds which could generate additional instabilities in the accretion disk \citep{2016AN....337..368D}. 

These cycles of variable accretion rates manifest themselves differently in the so-called ``Atoll" and ``Z" sources.  ``Z\index{Z}" sources show three X-ray spectral states, have higher accretion rates and tend to have longer orbital periods than ``Atoll\index{Atoll}" systems. The latter are driven by lower accretion rates and characterised by X-ray thermonuclear flashes (see. Sect. \ref{msp}).
 
\section{Accretion Column In Highly Magnetized Systems}

In highly magnetized systems, the plasma approaching the neutron star is stopped by the pressure of the dipolar pulsar magnetic field, independently of the way it flew from the companion star. The plasma is then forced to move along the field lines toward the magnetic poles, where it releases its gravitational energy in the form of high-energy radiation. This radiation is not emitted isotropically; misalignment of magnetic and rotational axes of the neutron stars ensure that periodic pulses of high-energy radiation could be detected.

\begin{figure}
\centerline{\includegraphics[width=8cm]{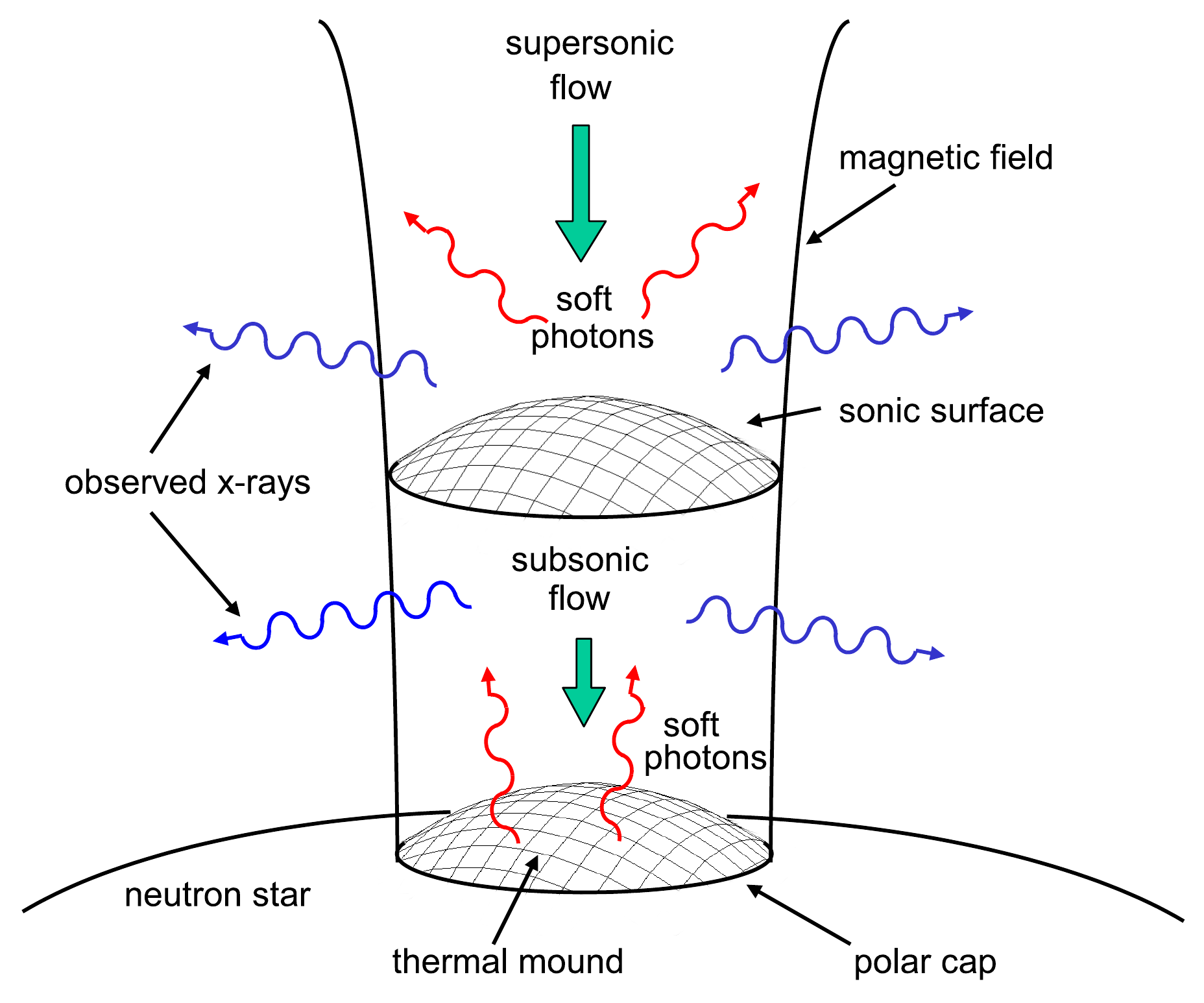}}
\caption{Simplified representation of an accretion column, where the X-ray radiation is produced in a high-magnetic field environment \citep[from][]{2007ApJ...654..435B}.}
\label{fig:column}
\end{figure}

Close to the neutron star surface, the plasma falls in a quasi-cylindrical accretion column\index{accretion column}, at a fraction of the speed of light (see a schematic representation in Fig.~\ref{fig:column}). It then heats to $10^8$ K \citep{1976MNRAS.175..395B,1985ApJ...299..138M}. Bulk and thermal Comptonization of seed photons produced by modified bremsstrahlung in the high magnetic field and black body emission from the column's base play a key role in the formation of the non thermal hard X-ray emission \citep{2007ApJ...654..435B}.

The X-ray continuum of accreting pulsars is characterized by a power law N$_\nu \sim \nu^{-(1-2)}$ with a high-energy exponential cutoff \citep[7-30 keV,][]{1983ApJ...270..711W}, sometimes modified by absorption and emission lines in the soft X-rays and by cyclotron resonance scattering features\index{cyclotron resonance scattering features} (CRSF) at higher energies \citep{1978ApJ...219L.105T}. CRSFs are caused by the scattering of hard X-ray photons on electrons whose energy is quantized by the magnetic field according to the Landau levels\index{Landau level}. Their energy separation can be measured from the source spectra and hence the magnetic field strength in the scattering region can be estimated. Variability of the CRSF with luminosity on long and spin period time scales indicate that the accretion flow is not uniform nor stationary \citep{1998AdSpR..22..987M,2016A&A...591A..29F}.

Observing transient X-ray pulsars in bright outburst (especially in Be systems) is essential to understand the physical processes at play close to the neutron star surface and in particular the response of the neutron star -- magnetosphere system to the variability of the mass accretion rate on different time scales. 

Modelling the interaction of the radiation with the accreted matter in strong magnetic and gravitational fields is a complex problem. A number of authors attempted to simulate the shape of the continuum \citep{2007ApJ...654..435B} and of the CRSFs \citep{2000ApJ...544.1067A, 2007A&A...472..353S} as a function of the pulse phase, source luminosity, geometry of the emission regions, etc. The comparison of the model predictions with the observations still fails to provide strong constrains on the physical parameters of the accretion column because of the complexity of the models, the limitations of current hard X-ray telescopes, and the convolution of the signatures of many emitting and absorbing regions with different properties.

The discovery of an anti-correlation between the cyclotron line energy and the X-ray luminosity in transient X-ray pulsars \citep{Mowlavi2006,2006MNRAS.371...19T} initiated a systematic study of the cyclotron line energy properties as a function of the source luminosity and was interpreted with a change of the geometry of the accretion column, rising above the neutron star surface at high luminosities \citep{Becker2012}. \citet{2014ApJ...781...30N} modelled the cyclotron line by the sum of the contributions emerging from individual line-forming regions along the accretion column with different magnetic field strength, temperature, and density. An increase of the mass accretion rate causes the emergence of additional line-forming regions with lower magnetic fields that lead to a decrease of the cyclotron line energy.
Another model \citep {2013ApJ...777..115P} suggests that a significant part of the accretion column radiation is intercepted and reflected by the neutron star surface because of relativistic beaming\index{beaming}. Variations of the accretion column\index{accretion column} height lead to a shift of the illuminated part of the neutron star surface toward the equator where the magnetic field is weaker. This naturally drives the observed anti-correlation of the cyclotron line energy with luminosity. Moreover, this model is able to explain why the amplitude of the cyclotron energy variability remains limited, when the luminosity changes dramatically.

For lower-luminosity sources an opposite behaviour of the cyclotron energy with the luminosity has been observed \citep[e.g.][]{Staubert2007,2012A&A...542L..28K}. This has been explained as due to the redshift of the line centroid energy due to the motion of the in-falling plasma: at low luminosity, the plasma is in free-fall, while  at higher luminosity the plasma slows down near the stellar surface, resulting in a reduced red-shift of the line centroid energy \citep{2015MNRAS.447.1847M}. Finally, no dependence of the cyclotron energy on luminosity has been detected for some transient pulsars \citep{2013ApJ...764L..23C}. Further observations of X-ray pulsars during bright outbursts are needed to discriminate between models.

\section{Low Magnetized Systems And Accreting Millisecond Pulsars\label{msp}}

There is a class of millisecond radio pulsars\index{millisecond radio pulsar} (MSPs) with about 300 members that have periods of rotation lower than $\sim$10\,ms \citep{Manchester2005}\footnote{\url{http://www.atnf.csiro.au/people/pulsar/psrcat/}}, while they slow down at an almost imperceptible rate. Since radiation is produced by emission of electromagnetic energy at the expense of their kinetic rotational energy, the product of period derivative and period is proportional to the pulsars's magnetic field. This implies that these objects have a relatively low surface magnetic field ($\sim10^8$\,G)  decayed by the higher typical initial value. therefore, they are old (Gyrs), but they must be spun up during their existence. To explain their origin, it was suggested that they are recycled pulsars\index{recycled pulsar} and that the spin-up occurred during a Gyr-long phase of accretion of mass transferred by a low-mass companion star through an accretion disk (e.g., \citealt{Bisnovatyi-KoganKomberg1974,Alpar1982,Radhakrishnan1982}). During the mass accretion phase these systems should be observed as bright LMXBs. In this phase, it has been argued 
that the magnetic field decays more rapidly than in an isolated neutron star, due to matter accumulated on the NS surface \citep[see ][and references therein]{Zhang2006}. When mass transfer declines, a pulsar powered by the rotation of its magnetic field turns on and shines mostly in the radio band.  

The very existence of millisecond pulsars powered by accretion was proven almost two decades after \citep{Wijnands1998} and since then less than 20 objects have been ascribed to the class of accreting millisecond pulsars (AMSP). These show only sporadic, month-long outburst during which the companion star overfills its Roche lobe, initiating a months-long mass transfer phase. An accretion disc is formed, which truncates at several neutron star radii, where the pressure of the pulsar's magnetic field equals the ram pressure of the infalling flow. In this phase, pulsars  are detected with an X-ray luminosity of  $\sim10^{37}$\,erg/s. For the rest of the time, they remain in quiescence, powered by rotation \citep{Burderi2003}. However, no observational evidence of that transition was found, until the discovery of the first object transiting back and forth between the rotation- and accretion-powered\index{accretion-powered} phases \citep[IGR\,J19245$-$2452;][]{Papitto2013}. In addition to the radio and X-ray bright states, it has been evidenced that this object presents variability patterns, which are significantly different from the other LMXBs, while it also exhibits levels of activity at an intermediate level ($10^{35}$\, erg/s) between the outburst and the X-ray quiescence ($10^{32}$\,erg/s).

At the time of writing, only two other objects have been found to switch between rotation-powered\index{rotation-powered} phase and this intermediate level of activity: PSR\,J1023+0038 \citep{Archibald2009} and XSS\,J1227.0$-$4859 \citep{Stappers2014}. Pulsations for these objects have been detected in the radio band when their X-ray luminosity is at the quiescent level and they were not surrounded by an accretion disk. The spin period modulations show that they are in a binary system with a low-mass companion and that there are frequent disappearances of the signal due to the shielding by a thick intra-binary medium, produced by the companion's evaporation\index{evaporation} due to the pulsar's wind\index{pulsar wind}. These systems can quickly (on a time scale of months) change their status when some mechanism triggers the formation of an accretion disc around the neutron star. In such phase, they shine in $\gamma$-rays\index{$\gamma$-ray} (0.03--300\,GeV) and increase their X-ray luminosity to an average value of $\sim10^{34}$\,erg/s (see also Tanaka, 2016). X-ray pulsations have been found while the radio pulsated signal is absent \citep{Archibald2015}. This has been interpreted as accretion in a strong propeller regime\index{propeller} with most of the matter ejected by the centrifugal motion of the pulsars magnetic field and some still reaching the surface to produce periodically modulated X-rays \citep{Papitto2015}.

X-ray pulsations from most of the other bright low mass X-ray binaries have eluded any detection with the exception of a handful of sources showing a coherent oscillation at the onset of their thermonuclear bursts\index{thermonuclear burst} \citep[see references in][]{Papitto2014}. On the contrary, when the source is powered by accretion, the pressure of the accreting material is so strong that the accretion disk likely extends down to the surface, inhibiting pulsations for symmetry reasons.

Thermonuclear bursts originate in weakly magnetised pulsars\index{weakly magnetised pulsar} because plasma deposit at the surface of the pulsar over a relatively large area without reaching the temperature necessary to ignite thermonuclear reactions. When enough material is accumulated, it can ignite explosive burning\index{explosive burning} (like in the core of normal stars), generating a powerful outburst with a duration of some tens or hundreds of seconds ($L_X\gg 10^{38}$\,erg/s). It is believed that the flames propagate through the atmosphere of the neutron star and, in the initial seconds, not all the surface is covered. Oscillations at the onset and tail of X-ray flashes\index{X-ray flash} allow us to detect the rotation of the star and to study the accretion and spreading of the plasma over the neutron star surface. On the contrary, for higher magnetic fields, the geometrical area over which the plasma is accreted is smaller, leading to conditions in which stable burning occurs.

To summarize, rotating neutron stars in close binary systems are believed to posses four main states with respect to accretion. They do not accrete and some of them shine as radio pulsars while the X-ray luminosity is $\sim10^{32}$\,erg/s; they accrete at a level in which the pressure of the accretion disk is almost entirely balanced by the magnetic pressure and produce strong outflows (L$_X\sim10^{34}$\,erg/s); the accretion disc is truncated at several neutron star radii and coherent pulsations are observed during weeks-long outbursts (L$_X\sim10^{37}$\,erg/s); some other systems possess very high accretion rates, however coherent pulsations are not observed excepting during some thermonuclear bursts.

Observing pulse profile of accretion-powered msec pulsars and of thermonuclear bursts with a high throughput X-ray instrument will allow to constrain the neutron star mass and radius with an accuracy of a few \% and to determine the dense matter equation of state\index{equation of state} \citep[Haensel and Zdunik, 2016;][]{2013ApJ...776...19L}.

\section{Summary}

About 500 pulsars in the Milky Way are known to have companion stars and many more systems remain undetected. Accretion-powered X-ray emission has been detected in about 250 of them, revealing a wealth of phenomena linked to plasma physics, magneto-hydrodynamics and general relativity. The accretion flow is driven by the nature of the stellar companion (high or low mass star), by the possible overflow of the Roche lobe and by the strength of the neutron star magnetic field.

The two classes of low- and high-mass X-ray binaries account for the main phenomenology. Future X-ray missions will allow to test fundamental physics in these systems, where extreme conditions of gravity, density and magnetic fields are reached.

\section*{Cross-References}
\begin{itemize}
\item X-ray Binaries (Casares and Israelian, 2016) 
\item Thermal Evolution of Neutron Stars (Geppert, 2016) 
\item Nuclear Matter in Neutron Stars (Haensel and Zdunik, 2016) 
\item Gamma Ray Pulsars; from Radio to Gamma Ray (Tanaka, 2016) 
\item Supernovae and the Evolution of Close Binary Systems (Van den Heuvel, 2016) 
\item Strange Quark Matter Inside Neutron Stars (Weber, 2016) 
\item Neutron Stars as Probes for General Relativity and Gravitational Waves (Wex, 2016)
\item Magnetic Field Evolution of Neutron Stars (Zhang, 2016) 
\end{itemize}

\bibliographystyle{spbasic}
\bibliography{reference} 

\end{document}